# Low-frequency gravitational waves from supermassive black holes

Martin G. Haehnelt

*Institute of Astronomy, Madingley Road, Cambridge CB3 0HA*
*Max-Planck-Institut für Astrophysik, Karl-Schwarzschild-Straße 1, D-85748 Garching*
*e-mail: de.mpg.mpa-garching@haehnelt*

## ABSTRACT

*Supermassive black holes are investigated as possible sources for low-frequency bursts of gravity waves. The event rate for 'known' supermassive black holes at intermediate and high redshifts, inferred from the quasar luminosity function is low $\sim 0.1\,\mathrm{yr}^{-1}$. A number density of gravitational-wave sources comparable to the number density of galaxies inferred from faint galaxy counts is necessary to raise the event rate significantly above one per year. A space-based interferometer could therefore only see several events per year from supermassive black holes if an additional population of supermassive black holes existed and emitted gravitational waves efficiently. These might reside in the population of dwarf galaxies or in a transient population of small dark matter haloes that have mostly merged into larger haloes hosting the galaxies seen today, as proposed in a hierarchical cosmogony. In the latter case, event rates could be of order $10 - 1000$ per year due to coalescing supermassive-black-hole binaries formed during the merging process. Event rates could be as high as a few per second if the dark matter in galaxy haloes consists of supermassive black holes in the mass range $10^3 - 10^6\,M_\odot$, and produced gravitational waves efficiently. The proposed space-based gravitational-wave interferometer LISA/SAGITTARIUS should detect most gravitational-wave events involving supermassive black holes above $10^4\,M_\odot$ out to redshifts of $z \sim 100$.*

## 1 INTRODUCTION

Bursts from supermassive black holes are — apart from periodic gravitational waves from known binary stars — the major possible source of low-frequency gravitational waves in the frequency range $10^{-5} - 1\,\mathrm{Hz}$. A considerable amount of work has been done on determining the strength, polarization and waveform of gravitational-wave bursts and on extracting these signals from noisy data (Peters & Mathews 1963; Estabrook & Wahlquist 1975; Armstrong, Estabrook & Wahlquist 1987, Thorne 1987; Tinto & Armstrong 1991 ; Schutz 1992; Thorne 1993), but little attention has been paid to the possible event rates since the first study of Thorne & Braginsky (1976).

The existence of supermassive black holes is still not proven beyond doubt, but a population of supermassive black holes in the mass range of $10^6 - 10^9\,M_\odot$ is believed to be the only viable explanation for most of the observed activity in active galactic nuclei (Lynden-Bell 1969, Blandford & Rees 1992). Thorne & Braginsky found an event rate $\sim 0.1\,\mathrm{yr}^{-1}$ for a source density comparable to the space density of typical $L_*$ galaxies (as defined by a fit of a Schechter-function to the galaxy luminosity function) and speculated that the observed number density of 'active' galaxies indicating the existence of supermassive black holes might be considerable higher than that. Since Thorne & Braginsky's paper the knowledge of the luminosity function of quasars and galaxies and of the build up of galactic-size structures has greatly improved. This renders a more detailed investigation possible.

Quasar activity seems to be quite common at intermediate and high redshift and probably most of the luminous galaxies have undergone a phase of quasar activity, but, despite this, the integrated number density of observed quasars is unlikely to exceed the number density of $L_*$ galaxies, even if the life time of the individual quasar is assumed to be rather short (Haehnelt & Rees 1993). A considerably higher number of supermassive black holes of more moderate mass ($10^4 - 10^6\,M_\odot$) could be contained in observed dwarf galaxies or in the building blocks of luminous galaxies which, in hierarchical cosmogonies, would be built up by merging of smaller units at moderate redshifts between $z = 4$ and $z = 0.5$ (White & Rees 1978). The large number of faint blue galaxies might be a direct observational indication for such a hierarchical build-up of structure involving high merging rates and probably a higher number of small galaxies than inferred from the local galaxy luminosity function (Tyson 1988, Guiderdoni & Rocca-Volmerange 1990, Kauffmann, White & Guiderdoni 1993). If the dynamically inferred dark matter in galactic haloes consists of a population of black holes, the fraction of all mass contained in black holes would be even a factor $10^6$ higher than that inferred for the black hole population powering quasars.

The second important question is whether a population of supermassive black holes is likely to emit detectable gravitational waves.

In principle, a significant fraction of the rest mass energy could be released during the formation of the black hole if it involves a non-axisymmetric collapse or rotation phase, but the experience from calculations for stellar mass size sources shows that the efficiencies are likely to be small and in the case of continuous slow accretion they could be even negligible (see Schutz 1989 for a review).

One source of gravitational radiation would be a close binary containing two supermassive black holes. It seems possible that supermassive black holes are formed as binaries by fission of a bar (Rees 1984) but a more likely path for the formation of a supermassive-black-hole binary is via the merging of the nuclei of two dark-matter haloes each containing a black hole. The putative black holes would sink to the centre of the merged nuclei on a dynamical friction time scale forming an initially rather wide binary. The newly-formed nucleus would be gas-rich and accretion onto the binary could lead to a rapid reduction of the separation. Effective emission of gravitational radiation would finally lead to coalescence (Begelman, Blandford & Rees 1980).

Another possibility for the emission of gravitational waves by a supermassive black hole is a head-on collision with a compact stellar object or the capture of such an object into a relativistic orbit.

Section 2 gives an analysis of which kind of gravitational-wave events involving supermassive black holes could be seen by a space-based gravitational-wave interferometer using the specifications of two recent proposals (LISA/SAGITTARIUS) for such a laser interferometer discussed in section 3 (Danzmann et al. 1993, Hellings et al. 1993). Section 4 discusses the expected event rates for different scenarios and section 5 contains the conclusions. A Hubble constant of $H_0 = 50\,\mathrm{km\,s^{-1}\,Mpc^{-1}}$ is used throughout the paper.

## 2 SUPERMASSIVE BLACK HOLES AS SOURCES OF LOW-FREQUENCY GRAVITATIONAL WAVES

The gravitational-wave amplitude produced by a burst releasing a fraction $\epsilon$ of the rest-mass energy of an emitting object of total mass $M$ is given by

$$h_{\mathrm{burst}} = \left( \frac{3}{2\pi^2} \frac{G\,(\epsilon M c^2)}{c^3\,R(z)^2\,f_{\mathrm{c}}} \right)^{1/2} \qquad (1)$$

(Thorne 1987), where

$$R(z) = \frac{c}{H_0} \frac{1 - q_0 + q_0 z - (1 - q_0)(1 + 2 q_0 z)^{1/2}}{q_0^2 (1 + z)} \qquad (2)$$

is the comoving coordinate distance as a function of redshift $z$. $q_0 = \Omega/2$ is the conventional deceleration parameter and equation (2) holds for a vanishing cosmological constant, $\Lambda = 0$. For an Einstein-de Sitter universe of critical density, $\Omega = \rho/\rho_{\mathrm{crit}} = 1$, equation (1) becomes

$$h_{\text{burst}} = 1.1 \times 10^{-20} \, h_{50} \epsilon_{0.1}^{1/2} \left( \frac{M}{10^4 \, M_\odot} \right) (1 - (1+z)^{-1/2})^{-1}, \tag{3}$$

where $h_{50}$ denotes the Hubble constant in units of $50 \, \text{km} \, \text{s}^{-1} \, \text{Mpc}^{-1}$. The characteristic burst frequency is related to the mass of the black hole and the redshift at which the burst occurs as

$$f_c = [(3^{3/2} GM/c^3)(1+z)]^{-1} = 3.9 \left( \frac{M}{10^4 \, M_\odot} \right)^{-1} (1+z)^{-1} \, \text{Hz} \tag{4}$$

(Press 1971). For a binary with component masses $M_1$ and $M_2$, the amplitude is given by

$$\begin{aligned}
h_{\text{period}} &= 8 \left( \frac{2}{15} \right)^{1/2} \frac{G^{5/3} \mu}{c^4 \, R(z)} (\pi M f)^{2/3} \\
&= 3.4 \times 10^{-22} \, h_{50} \left( \frac{f}{10^{-3} \, \text{Hz}} \right)^{2/3} \left( \frac{M_1}{10^4 \, M_\odot} \right)^{5/3} \frac{x}{(1+x)^{1/3}} (1 - (1+z)^{-1/2})^{-1}
\end{aligned} \tag{5}$$

(Thorne 1987, Danzmann et al. 1993), where $M = M_1 + M_2$ and $\mu = M_1 M_2 / M$ are the total and the reduced mass and $x = M_2/M_1$ denotes the mass ratio.

The detector sensitivity as a function of frequency determines the most and the least massive black hole (binary) for which the detection of a burst is possible. This depends of course on redshift, emission efficiency and, in case of a binary, on the mass ratio. The proposed space based laser interferometer LISA/SAGITARRIUS (see section 3 for details of the sensitivity limits) would see gravitational wave bursts in the wedge shaped area above and to the left of the thick solid curves in Fig. 1. This assumes $\epsilon = 0.1$ as might be realistic for the coalescence of a binary with equal masses, and the thin solid curves show the limits for $\epsilon = 10^{-3}$ and $\epsilon = 10^{-5}$. For binaries of unequal masses the sensitivity is reduced by a factor $x^{-1/2}$ at the frequency of coalescence and the thin solid curves can therefore also be interpreted as the limits for the coalescence of binaries with mass ratios $x = 10^{-2}$, $x = 10^{-4}$ (this assumes that the form of the signal is known with sufficient accuracy to apply matched filtering). For the coalescence of binaries of small mass occurring at small redshift, the highest sensitivity is obtained long before coalescence due to the narrow range in frequency where the interferometer reaches its peak sensitivity. The thick dashed curve shows the lower mass limit for a 'wide' binary with equal masses and the thin dashed curves show the sensitivity for binaries with mass ratios $x = 10^{-2}, 10^{-4}$. Here the signal is assumed to be integrated for the last year before coalescence.

A gravitational-wave detector with the sensitivity of LISA/SAGITTARIUS would see most possible gravitational-wave events involving supermassive black holes above $10^4\,M_\odot$ back to very high redshifts, thus maximizing the possible event rates. A detector with a somewhat lower sensitivity would still cover a considerable fraction of the $(M,z)$ plane, but as will be shown in section 4, even a sampling out to high redshifts does not necessarily lead to event rates sufficient to guarantee a few events per year.

Compact objects of $1\,M_\odot$ in a relativistic orbit around a black hole of $10^6\,M_\odot$ would be seen out to moderate redshifts $z \sim 1-3$. Events involving black holes of mass $\lesssim 10^3\,M_\odot$ are only seen at non-cosmological distances and a binary of stellar mass black holes ($10\,M_\odot$) at the distance of Virgo might be just visible.

# 3 THE PROPERTIES OF LISA/SAGITTARIUS

LISA (Laser Interferometer Space Antenna) is a proposed space-based Michelson laser interferometer in orbit around the Sun designed to detect low-frequency gravitational waves in the frequency range $10^{-5} - 1\,\text{Hz}$ not detectable on Earth due to unavoidable seismic noise (Danzmann et al. 1993). Due to the vastly increased armlength ($5 \times 10^6$ km) compared to a ground based interferometer like LIGO, the comparatively moderate accuracy in the determination of changes in the arm length achievable in space is sufficient to reach a sensitivity similar to LIGO in its advanced form. The SAGITTARIUS project tries to reach the same sensitivity specifications with a space-based interferometer orbiting the Earth (Hellings et al. 1993). The armlength is similar ($10^6$ km) and the projected residual spurious accelerations and position measuring noise are also comparable to the LISA project.

The sensitivity will be limited by spurious accelerations at frequencies below $10^{-3}$ Hz and by shot noise at higher frequencies. For randomly oriented sources a search for gravitational bursts in a year of data with a Wiener filter renders a sensitivity of

$$h_{\text{burst}} \sim 10 S_h f_c^{1/2} \qquad (6)$$

(Thorne 1987), where $S_h$ is the spectral noise density of the detector (in $\text{Hz}^{-1/2}$) and $f_c$ is the characteristic frequency of the burst. A corresponding one-year search for periodic sources gives

$$h_{\text{periodic}} \sim 10 S_h (3.15 \times 10^7\,\text{s})^{-1/2}, \qquad (7)$$

for a signal-to-noise ratio $S/N = 5$. The projected maximum spurious acceleration for LISA is $3 \times 10^{-13} \, (f/10^{-4} \, \text{Hz})^{-1/3} \, \text{cm s}^{-2} \, \text{Hz}^{-1/2}$, and the planned limit for the accuracy of the armlength measurement due to frequency-independent shot noise is $2 \times 10^{-9} \, \text{cm Hz}^{-1/2}$. For an armlength of $5 \times 10^6$ km, the sensitivity for bursts is then approximately given by

$$h_{\text{burst}} \sim \begin{cases} 1.3 \times 10^{-21}(f/10^{-3} \, \text{Hz})^{-11/6}, & f < 10^{-3} \, \text{Hz}, \\ 1.3 \times 10^{-21}(f/10^{-3} \, \text{Hz})^{1/2}, & 10^{-3} \, \text{Hz} \le f \le 3 \times 10^{-2} \, \text{Hz}, \\ 6.9 \times 10^{-21}(f/3 \times 10^{-2} \, \text{Hz})^{3/2}, & f > 3 \times 10^{-2} \, \text{Hz}, \end{cases} \quad (8)$$

The break at $3 \times 10^{-2}$ Hz is due to the fact that there $\lambda/2$ equals the armlength of the interferometer. The corresponding sensitivity for periodic sources is

$$h_{\text{periodic}} \sim \begin{cases} 7.1 \times 10^{-24}(f/10^{-3} \, \text{Hz})^{-7/3}, & f < 10^{-3} \, \text{Hz}, \\ 7.1 \times 10^{-24}, & 10^{-3} \, \text{Hz} \le f \le 3 \times 10^{-2} \, \text{Hz}, \\ 7.1 \times 10^{-24}(f/3 \times 10^{-2} \, \text{Hz},) & f > 3 \times 10^{-2} \, \text{Hz}. \end{cases} \quad (9)$$

The sensitivity limits for SAGITTARIUS are almost identical.

## 4 EVENT RATES FOR GRAVITATIONAL WAVES FROM SUPERMASSIVE BLACK HOLES

For a population of gravitational-wave sources with comoving density $N_{\text{s}}(z)$, emitting at a rate $\nu_{\text{em}}$, the total rate of received gravitational-wave events originating around redshift $z$ between comoving coordinate distance $R(z)$ and $R(z) + \text{d}R$ is given by

$$\text{d}\nu_{\text{rec}}(z) = \nu_{\text{em}}(z) \, (1+z)^{-1} \, N_{\text{s}}(z) \, 4\pi \, R^2(z) \, \frac{\text{d}R}{\sqrt{1 - k \, (R(z)/\mathcal{R})^2}}, \quad (10)$$

where $k = -1, 0, 1$ for an open, flat or closed universe respectively, and $\mathcal{R} = c/(H_0 \, (1 - \Omega - \Lambda/3H_0^2))$. For an emission rate of one per time interval corresponding to a unit increment in redshift
$\nu_{em} = H_0 \, (1+z) \, (\Omega \, (1+z)^3 + \Lambda/3H_0^2)^{1/2}$ and with the transformation from comoving coordinate interval to redshift interval
$\text{d}R/\sqrt{1 - k \, (R(z)/\mathcal{R})^2} = c \, H_0^{-1} \, (\Omega \, (1+z)^3 + \Lambda/3H_0^2)^{-1/2} \, \text{d}z$ equation (10) becomes

$$\frac{\mathrm{d}\nu_{\mathrm{rec}}(z)}{\mathrm{d}z} = 4\pi \ R(z)^2 \ N_{\mathrm{s}}(z) \ c. \qquad (11)$$

For an Einstein-de Sitter universe this becomes

$$\frac{\mathrm{d}\nu_{\mathrm{rec}}(z)}{\mathrm{d}z} = 5.6 \ h_{50}^{-2} \left(\frac{N_{\mathrm{s}}(z)}{10^{-2} \ \mathrm{Mpc}^{-3}}\right) \ (1 - (1+z)^{-1/2})^2 \ \mathrm{yr}^{-1}. \qquad (12)$$

In the following I will assess the space density of possible sources and the corresponding event rates for an Einstein-de Sitter universe.

### 4.1 Observed quasars as sources of gravitational waves

Due to the large surveys for high redshift quasars undertaken in recent years, the quasar luminosity function and its evolution with redshift has been determined quite accurately out to a redshift $z \sim 4 - 4.5$ (Boyle et al. 1988; Boyle et al. 1991; Boyle 1993; Irwin, McMahon & Hazard 1991; Warren, Hewett & Osmer 1991; Warren, Hewett & Osmer 1993). With the assumption that quasars are short-lived (life time $t_{\mathrm{Q}} \sim 10^8$ yr) and that each of the black holes produces a gravitational wave event at the redshift where the quasar is observed (e.g. by coalescing of two black holes during, shortly before or shortly after the observed activity) the number density of observed quasars in a given luminosity range ($L \ \Phi(L, z)$) can be transformed into a corresponding number density of supermassive black holes emitting once per unit increment in redshift

$$\frac{\mathrm{d}N_s(M, z)}{\mathrm{d}\ln M} = \frac{1}{t_{\mathrm{Q}} \ H_0 \ (1+z)^{5/2}} \ L \ \Phi(L = \delta M). \qquad (13)$$

$\Phi(L) \, \mathrm{d}L$ is the number density of quasars in the luminosity range $\mathrm{d}L$ and a linear transformation of quasar luminosity to black hole mass $\delta$ is assumed. The numerical value of $\delta$ is somewhat uncertain but is of secondary interest for an estimate of the event rates as a LISA-like detector would be sensitive to all events out to a redshift $z \sim 100$ for the mass range in question. Long- and short-lived models for the quasar population are discussed in the literature (e.g. Cavaliere et al. 1983), but a detailed examination of the quasar luminosity function and its evolution (Haehnelt & Rees 1993) seems to indicate a rather short lifetime of order of the Salpeter time $t_{\mathrm{Salp}} \sim 4 \, \epsilon_{\mathrm{rad},0.1} \, 10^7$ yr (the e-folding time for the mass of a black hole accreting at the Eddington limit and radiating with efficiency $\epsilon_{\mathrm{rad}}$). Fig. 2 shows the expected event rates for a given mass range as obtained from the quasar luminosity function integrated from redshift $z = 0$ to redshift $z_1$, where $z_1$ varies in steps of 0.5 (solid curves). The event rates would be dominated by quasars at redshifts $z \leq 3$ where the luminous function is rather unanimously determined and the luminosity function of Boyle et al. (1991) is used. At higher redshift the luminosity function is somewhat controversial and in order to show that the assumed luminosity function at $z \geq 3$ has little effect on the total event rate, a constant space density is assumed for $z \geq 3$. This overestimates the number density of luminous quasars with $z \geq 3$ by a factor of up to 10 compared to the one found by Warren et al. (1993). The differential rate (per redshift interval) is shown by the dashed curves. It is clearly seen that the number density of black holes inferred from the quasar luminosity function can not generate more than one event every few years if each black hole produces only one event during its active lifetime. Even a smooth extrapolation of the quasar luminosity function to smaller masses does not raise the rate above one per year.

However, high redshift quasar surveys have a rather bright limiting absolute magnitude and nothing is known about the faint end of the quasar luminosity function. An additional more numerous population of massive black holes of $10^4 - 10^6\,M_\odot$ radiating at the Eddington limit would not have been detected in current surveys at redshifts larger than $0.5 - 1$. It could furthermore be that such a population does not radiate efficiently.

### 4.2 Faint/Dwarf galaxies as hosts of massive black holes

In this section the event rates are investigated for the case that a population of massive black holes ($10^4 - 10^6\,M_\odot$) resides in observed small galaxies and has produced gravitational waves efficiently at some redshift. To estimate the number density of sources, I first use the local galaxy luminosity function. Galaxy luminosity functions are well described by a Schechter function

$$\Phi\,\mathrm{d}L = \Phi_* \left(\frac{L}{L_*}\right)^\alpha \exp\left(-L/L_*\right) \frac{\mathrm{d}L}{L_*}, \tag{14}$$

were $\alpha$ describes the slope at the faint end and $\phi_*, L_*$ are the characteristic number density and luminosity at the bend in the function. Fig. 3 shows the event rate inferred from the number density of all galaxies more luminous than a given luminosity $L$ as a function of $L/L_*$ assuming

$$N_\mathrm{S}(L/L_*, z, z_1) = \begin{cases} \int_L^\infty \Phi(L)\,dL, & z_1 - 0.5 \lesssim z \lesssim z_1 + 0.5, \\ 0, & \text{otherwise}, \end{cases} \tag{15}$$

for three different values of $\alpha$ (-1.07,-1.25,-2.0) and $\Phi_* = 2 \times h_{50}^3 10^{-3}\,\mathrm{Mpc}^{-3}$. The thick solid curve ($\alpha = -1.07$) corresponds to the luminosity function obtained by Efstathiou et al. (1988) and a redshift for gravitational-wave emission $z_1 = 3.0$. All galaxies down to $0.01 L_*$ are required to raise the event rate to a few per year.

A considerably larger number density of small galaxies is inferred from deep CCD imaging (Tyson 1988). One interpretation of this apparent discrepancy between the faint galaxy counts and the faint end of the local galaxy luminosity function is a merging of these galaxies into larger galaxies (Guiderdoni & Rocca-Volmerange 1990, Guiderdoni & Rocca-Volmerange 1991, Kauffmann et al. 1994). With the assumption that each of these galaxies has emitted gravitational waves at redshift $z_1$ (by formation of a supermassive-black-hole binary), the number counts of faint galaxies can be converted into a number density of gravitational-wave sources

$$N_{\rm s}(\mathcal{N}, z_0, z_1, z) = \begin{cases} 4\pi \, \mathcal{N}/V(z_0) & z_1 - 0.5 \lesssim z \lesssim z_1 + 0.5, \\ 0, & \text{otherwise}, \end{cases} \qquad (16)$$

where $\mathcal{N}$ is the number of galaxies per steradian and $V(z_0)$ the integrated comoving volume out to the typical redshift of the galaxies $z_0$. Tyson et al (1988) find $\mathcal{N} = 4.5 \times 10^5 \, {\rm deg}^{-2}$ for a limiting magnitude $B = 27$ and $z_0 = 0.5$. The obtained event rate is then somewhat more favourable

$$\nu = 1.4 \, h_{50} \left( \frac{\mathcal{N}}{4.5 \times 10^5 {\rm deg}^{-2}} \right) \frac{(1 - (1 + z_1)^{-1/2})^2}{(1 - (1 + z_0)^{-1/2})^3} \, {\rm yr}^{-1}, \qquad (17)$$

and could be of order 30 per year (Fig. 4).

### 4.3 Merging in a hierarchical cosmogony as source of gravitational waves

The rather low possible event rate inferred for small galaxies is due to the flat nature of the local luminosity function at the faint end (for $\alpha = -1$ the number of objects per $\ln L^{-1}$ is constant). The faint counts already indicate that there might be more small structures, and it is a characteristic feature of hierarchical cosmogonies that they predict the number of haloes to rise roughly as $M^{\alpha_{\rm m}+1}$ per $\ln M^{-1}$ with $\alpha_{\rm m} \sim -1.8 - -1.9$. For a constant mass-to-light ratio this would correspond to a luminosity function with $\alpha = \alpha_{\rm m}$ close to $-2$ (see dotted curve in Fig. 3). In a hierarchical scenario, a specified dark matter halo of mass $M_1$ has typically undergone a merging history involving $\sim \beta M_1/M_0$ smaller haloes of mass $M_0$, where $1 - \beta$ indicates the fraction of mass accreted in the form of haloes less massive than $M_1$. Recent numerical and analytical studies of the merging of dark-matter haloes by Lacey & Cole (1993) indicate that $1 - \beta$ should rapidly decrease with increasing $M_1/M_0$, and I will consider $\beta = 0.3$ as a lower limit for big values of $M_1/M_0$. If all building blocks of mass $M_0$ contained a black hole and emitted gravitational waves at redshift $z_1$ the hierarchical build-up of galaxies brighter than $L_*$ would lead to an event rate

$$\nu \geq 14 \left(\frac{\beta}{0.3}\right) \left(\frac{\gamma_{\text{eff}}}{10^{-5}}\right) \left(\frac{M_{\text{min}}}{10^4 \, M_\odot}\right)^{-1} \, \text{yr}^{-1}. \qquad (18)$$

This assumes a mass of $5 \times 10^{11} \, M_\odot$ for an $L_*$ galaxy, a minimum black-hole mass $M_{\text{min}}$ which can be seen out to very high redshifts, a ratio $\gamma_{\text{eff}}$ of black hole mass to mass of the dark-matter haloes merging and an emission at $z_1 \geq 1$.

However, the emission of gravitational-wave events due to the merging of putative black holes within merging dark-matter haloes does not necessarily have to be coupled to the occurrence of luminous galaxies. Hierarchical cosmogonies generally predict many more (small) dark-matter haloes than indicated by the number of (faint) galaxies (Kauffmann et al. 1994). For a given hierarchical cosmogony — characterized by its power spectrum of primeval density fluctuations — the Press-Schechter formalism allows simple analytical estimates of the number density of haloes as a function of mass and redshift

$$\frac{\mathrm{d}n_{\text{halo}}}{\mathrm{d}\ln M}(z) = \left(\frac{2}{\pi}\right)^{1/2} \frac{\rho_0}{M} \left|\frac{d\ln \sigma}{d\ln M}\right| \frac{\delta_c}{\sigma(M,z)} \exp\left[\frac{-\delta_c^2}{2\sigma(M,z)^2}\right], \qquad (19)$$

where $\sigma(M,z)$ is the variance in the mass density $\rho_0$ and $\delta_c \sim 1.68$ is the linearly extrapolated overdensity at collapse (Press & Schechter 1974, Lacey & Cole 1993). Lacey & Cole give the merging rate of haloes of mass $M$ with haloes of the same mass or larger

$$\frac{\nu_{\text{merg}}(M,z)}{H_0 \, (1+z)^{3/2}} = \left(\frac{2}{\pi}\right)^{1/2} \int_M^\infty \frac{\delta_c}{\sigma(M+M',z)} \frac{1}{(1-\sigma^2(M+M',z)/\sigma^2(M,z))^{3/2}}$$
$$\left|\frac{d\ln \sigma}{d\ln (M+M')}\right| \exp\left[\frac{-\delta_c^2}{2}\left(\frac{1}{\sigma^2(M+M',z)} - \frac{1}{\sigma^2(M,z)}\right)\right] \frac{\mathrm{d}M'}{M+M'}. \qquad (20)$$

With the assumption that each merger leads to the formation of a supermassive-black-hole binary and subsequent emission of gravitational waves, this can be easily transformed into a number density of gravitational wave sources emitting once per unit increment in redshift

$$\frac{\mathrm{d}N_s(M,z)}{\mathrm{d}\ln M} = \frac{\mathrm{d}n_{\text{halo}}}{\mathrm{d}\ln M} \, \nu_{\text{merg}}(M,z) \, H_0^{-1}(1+z)^{-5/2}. \qquad (21)$$

The solid curves in figure 5 show the integrated event rates corresponding to mergers between redshift $z$ and the present for two specific hierarchical cosmogonies, a standard cold dark matter scenario (with variance in the mass density in $8h_{100}^{-1}$ Mpc spheres $\sigma_8 = 0.4$) and a mixed dark matter model (25 per cent neutrino fraction, COBE normalization), but the numbers depend only weakly on the details of the chosen scenario. For halos smaller than $10^{10}\,M_\odot$ the event rate exceeds 30 per year. If bursts from supermassive black holes with masses larger than $10^4\,M_\odot$ could be seen out to redshifts $z = 4$, a black hole formation efficiency as small as $\gamma_{\text{eff}} = M_{\text{black hole}}/M_{\text{halo}} \sim 10^{-6}$ could guarantee a generous event rate for a space-based interferometer. Should the black hole formation efficiency be higher and should therefore the merging of smaller haloes lead to the emission of detectable gravitational-wave bursts, the event rates could be considerably higher.

### 4.4 Black holes as dark matter in galaxy haloes

One candidate for the dynamically inferred dark-matter haloes of galaxies is massive black holes in the mass range $10^3 - 10^6\,M_\odot$ (Carr, Bond & Arnett 1984, Hut & Rees 1993, Rix & Lake 1993). This would enhance the number density of black holes by a factor of order $10^3 - 10^6$ compared to the estimates discussed in the previous sections. However, it is by no means clear that such a population of black holes would emit gravitational waves as the most likely mechanism — merging of the host objects and subsequent formation of a close binary— is missing. For the mass range in question, the event rate would only be sufficient if either the efficiency $\epsilon$ is high enough to allow detection at high redshift or if the whole population is synchronized to undergo a gravitational-wave burst at very low redshift (Bond & Carr 1984). In the first case the event rate is given by

$$\nu \geq 3.3 \times 10^7 \left(\frac{\Omega_{\text{bh}}}{0.1}\right) \left(\frac{M}{10^4}\right)^{-1} \text{yr}^{-1}, \quad (22)$$

where $\Omega_{\text{bh}}$ is the fraction of the critical density in black holes of mass $M$ and emission at $z_1 \geq 1$ is assumed. A small fraction of close binaries coalescing in a Hubble time would therefore be sufficient to reach an event rate above a few per year.

### 4.5 Influence of the cosmological parameter and the redshift of the gravitational wave bursts

For a given comoving number density $N(z)$ of sources emitting once around redshift $z_1$, the total received rate increases monotonically with growing $z_1$ due to the increasing comoving volume sampled. The event rates increase by a factor 4.1 between $z_1 = 1$ and $z_1 = 5$, and a further increase does not change the event rates considerably. For $z_1 = \infty$ the rate would be a factor 2.9 higher than at $z_1 = 5$. Lowering $\Omega$ and/or introducing a cosmological constant increases the comoving volume at a given redshift. For $\Omega = 0.2, \Lambda = 0$, the event rate would be higher by a factor (1.4,3.2,25.0) at $z = (1, 5, \infty)$ respectively and for $\Omega = 0.2, \lambda = \Lambda/3H_0^2 = 0.8$, the corresponding factors are (2.0,3.1,3.8). However it should be kept in mind that the bigger comoving volume also reduces the estimates of the comoving density of gravitational-wave sources inferred from high redshift objects.

# 5 CONCLUSIONS

While a detector with the sensitivity of LISA should certainly see gravitational radiation emitted from binary stars in our or nearby galaxies, there are several assumptions required in the case of gravitational wave bursts from supermassive black holes:

- Supermassive black holes have to exist.

- Supermassive black holes have to emit gravitational waves effectively either when they are formed, or as (coalescing) supermassive binaries.

- The number of gravitational-wave events involving supermassive black holes must be considerably higher than the one inferred from the number density of observed quasars. The number of sources has to approach or exceed the number of all (including the faintest) observed galaxies to guarantee event rates above one per year.

The first assumption is widely accepted to be true and the second and the third would be very plausible in a hierarchical cosmogony if black hole formation occurs in the large number of small dark matter halos that inevitably have to form and a considerably fraction of which has to merge into larger structures. Rather inefficient black-hole formation in these small haloes would be sufficient, and typical event rates are of order $10 - 1000\,\text{yr}^{-1}$.

The question of whether the formation of such a population of low-mass black holes is likely is difficult to assess. While it is certainly plausible that the gas in these haloes is able to cool and loose angular momentum (Loeb 1992), the potential wells would be rather shallow so that feedback processes of massive stars might prevent a concentration of the gas. On the other hand, the formation of one supermassive star of $10^4\,M_\odot$ collapsing into a black hole would be sufficient.

If the dynamically inferred dark matter haloes of galaxies consisted of black holes of $10^3 - 10^6\,M_\odot$ — a considerably more speculative assumption — , a rather small fraction of these holes producing gravitational waves efficiently (e.g. in binaries coalescing within a Hubble time) would guarantee a sufficient event rate. If all these black holes produced gravitational waves efficiently, the event rate could reach a few per second exceeding the confusion limit and producing a continuous background.

In summary, it can be said, that the population of supermassive black holes that a space-based gravitational-wave interferometer could see, has not yet been observed by other means. However, the existence of such a population is certainly plausible, and the fact that some speculation is involved might be outweighed by the fact that such a population could probably not be detected otherwise.

## ACKNOWLEDGMENTS

I would like to thank Philipp Podsiadlowski for comments on the manuscript and Martin Rees for suggesting this investigation and many helpful discussions. I acknowledge support by an Isaac Newton Studentship and the Gottlieb Daimler- and Karl Benz-Foundation.

# FIGURE CAPTIONS

**Fig. 1.** The LISA sensitivity limits for gravitational waves emitted from supermassive black hole (binaries) of total mass $M$ as a function of redshift. LISA is able to detect bursts with emission efficiency $\epsilon$ in the the wedge shaped area to the left and above the solid curves. The dashed curves show the reduced lower mass limit for coalescing binaries with mass ratio $x$. The dotted curve shows the detection limits for the spiral-in of a $1\,M_\odot$ compact object.

**Fig. 2.** The rate of gravitational-wave bursts produced by black holes of mass $M$ as inferred from the quasar luminosity function. For $z \leq 3$ the luminosity function determined by Boyle et al. (1991) is used with the assumption that each quasar produces one gravitational-wave event during a lifetime of $4 \times 10^7$ yr. For $3.0 \leq z \leq 4.5$ the luminosity function is assumed to be constant. Solid curves show the integrated event rate between $z = 0$ and $z = (0.5, 1.0, 1.5. 2.0, 2.5, 3.0, 3.5, 4.0, 4.5)$ respectively. Dashed curves show the differential event rate per corresponding redshift interval. Event rates scale inversely with the assumed quasar lifetime.

**Fig. 3.** The event rate inferred for galaxies more luminous than a certain fraction of $L_*$ from a Schechter-type galaxy luminosity function. It is assumed that each galaxy hosts a black hole capable of producing one detectable gravitational-wave event at redshift $z$. Solid curves are for the galaxy luminosity function as determined by Efstathiou et al. (1988) with $\alpha = -1.07$ and $\Phi_* = 2 \times 10^{-3} h_{50}^3\,\mathrm{Mpc}^{-3}$. The assumed emission redshift is $z = 3$ (thick curves) and $z = 1.0, 5.0$ (thin curves). The dashed curve is for $\alpha = -1.25$ and the dotted curve for $\alpha = -2.0$.

**Fig. 4.** The event rate for galaxies brighter than $B = 27$ inferred from faint counts (Tyson et al. 1988) with the assumption that the average galaxy in the faint counts has redshift $z_0$ and each galaxy hosts a black hole capable of producing a detectable gravitational-wave event at redshift $z_1$. The event rate scales linearly with the assumed normalization ($\mathcal{N} = 4.5 \times 10^5 \mathrm{deg}^{-2}$).

**Fig. 5.** The integrated event rates for dark-matter haloes of mass $M$ merging between redshift $z$ and the present for a standard cold-dark-matter scenario ($\sigma_8 = 0.4$) and a mixed-dark-matter model (neutrino fraction 25 per cent, COBE normalization). It is assumed that one gravitational-wave burst is emitted at the merging redshift.

Fig. 1

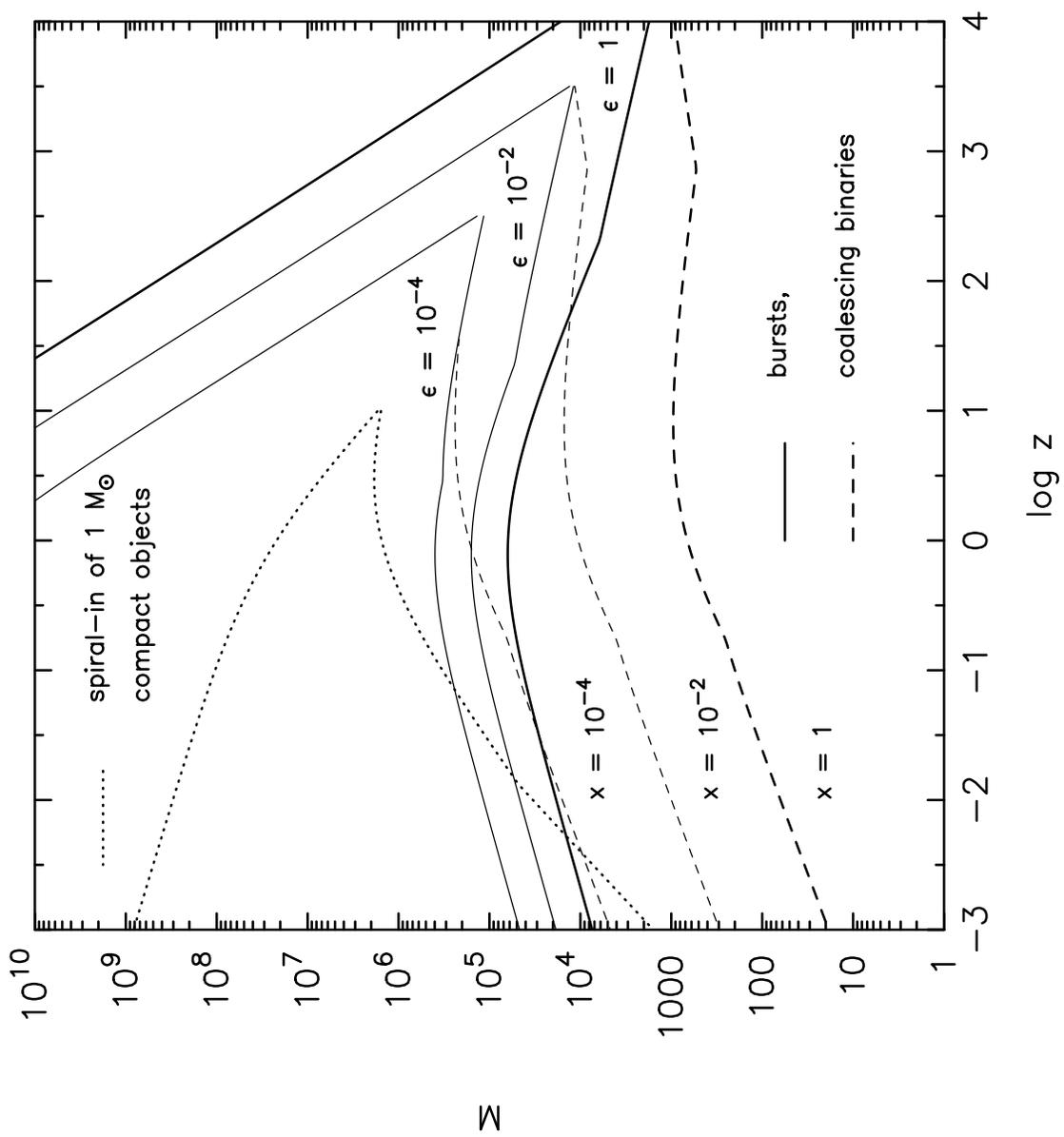

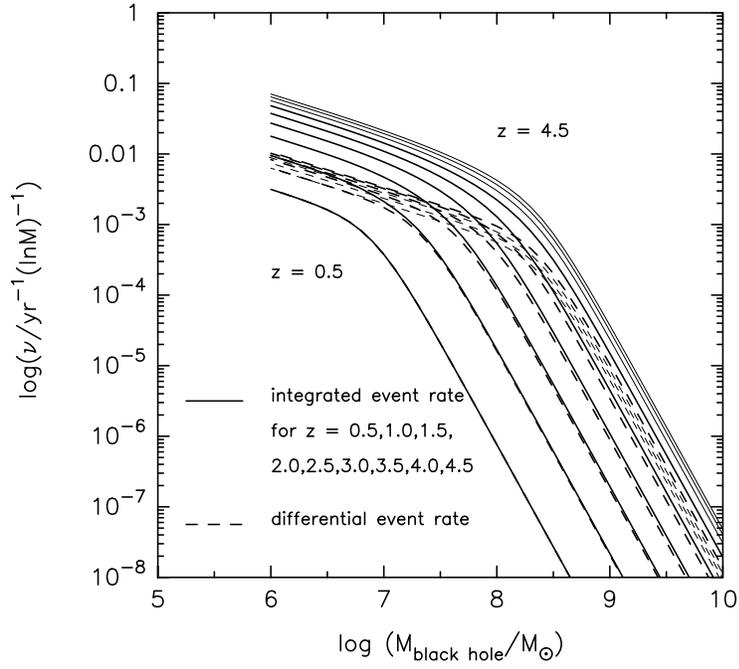

Fig. 2

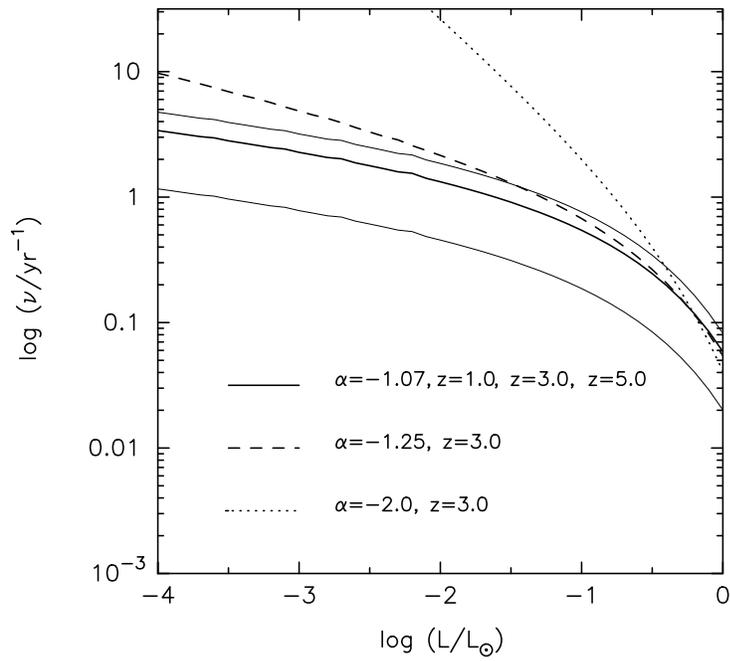

Fig. 3

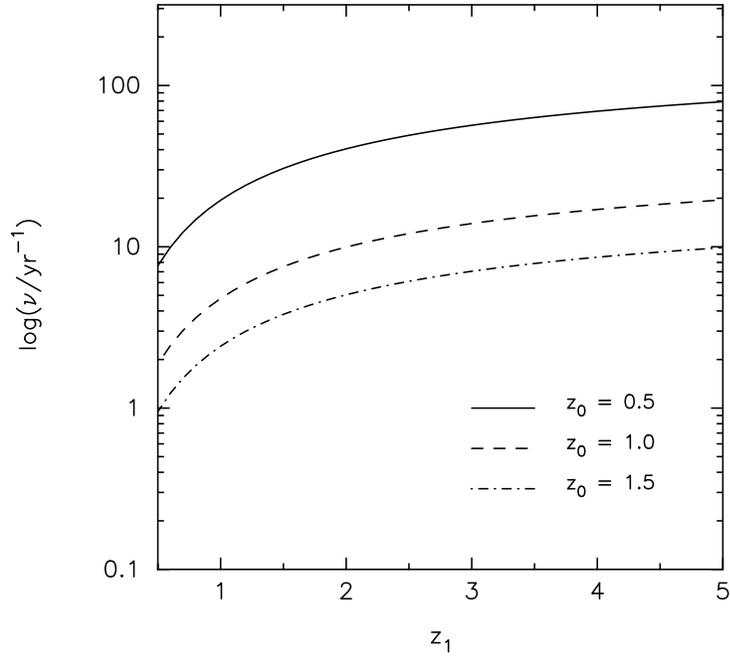

Fig. 4

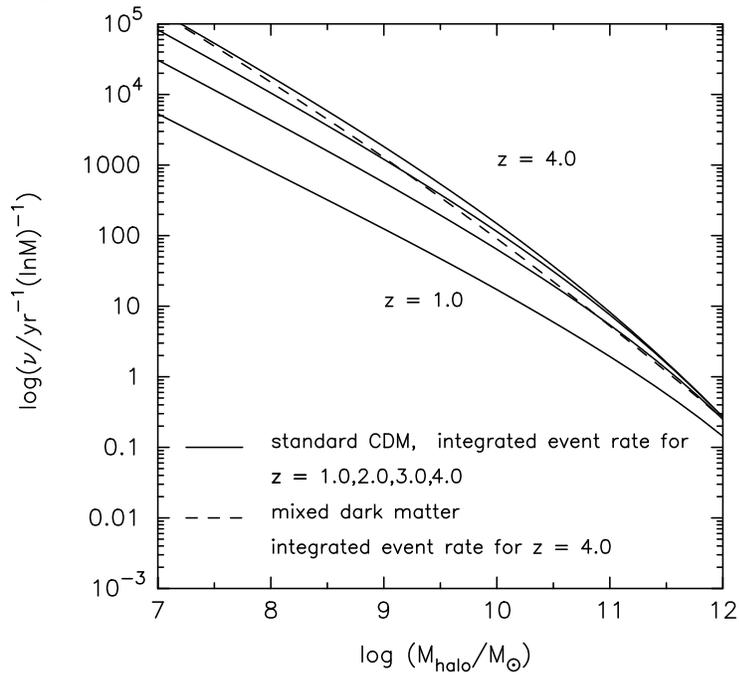

Fig. 5